\documentclass[aps,prb,twocolumn,superscriptaddress,epsfig,amsmath,showpacs]{revtex4}
\usepackage{epsfig}
\usepackage{dcolumn}
\usepackage{bm}
\begin{document}
\title{Prediction of superconducting properties of CaB$_2$ using anisotropic Eliashberg theory}

\author{Hyoung Joon Choi}
\email[Email:\ ]{h.j.choi@yonsei.ac.kr}
\affiliation{Department of Physics and IPAP, Yonsei University, 
Seoul 120-749, Korea}
\author{Steven G. Louie}
\affiliation{Department of Physics, University of California,
Berkeley, California 94720, USA}
\affiliation{Materials Sciences Division, Lawrence Berkeley National Laboratory,
Berkeley, California 94720, USA}
\author{Marvin L. Cohen}
\affiliation{Department of Physics, University of California,
Berkeley, California 94720, USA}
\affiliation{Materials Sciences Division, Lawrence Berkeley National Laboratory,
Berkeley, California 94720, USA}

\date{\today}

\begin{abstract}
Superconducting properties of hypothetical simple hexagonal CaB$_2$ are 
studied using the fully anisotropic Eliashberg formalism based on electronic 
and phononic structures and electron-phonon interactions which are obtained 
from {\em ab initio} pseudopotential density functional calculations.
The superconducting transition temperature $T_c$, the superconducting energy 
gap $\Delta({\bf k})$ on the Fermi surface, and the specific heat are 
obtained and compared with corresponding properties of MgB$_2$. Our results 
suggest that CaB$_2$ will have a higher $T_c$ and a stronger two-gap nature,
with a larger $\Delta({\bf k})$ in the $\sigma$ bands but a smaller 
$\Delta({\bf k})$ in the $\pi$ bands than MgB$_2$.

\end{abstract}


\maketitle

The discovery of superconductivity in MgB$_2$\cite{nagamatsu}
with its remarkable superconducting transition temperature $T_c$ as high as 39~K
has triggered many investigations of related materials including
various binaries: BeB$_2,$\cite{satta01,medvedeva01prb,ravindran01}
CaB$_2,$\cite{ medvedeva01jetp,okatov01,medvedeva01prb,ravindran01,oguchi02}
NaB$_2,$\cite{oguchi02}
SrB$_2,$\cite{ravindran01,oguchi02}
CuB$_2$,\cite{mehl01}
ZrB$_2$,\cite{gasparov01,oguchi02}
TaB$_2$,\cite{rosner01,oguchi02} 
OsB$_2$,\cite{singh07}
ScB$_2$,\cite{medvedeva01prb,oguchi02} 
YB$_2$,\cite{medvedeva01prb,oguchi02}
MgBe$_2$,\cite{mehl01} 
MgB$_4$,\cite{ravindran01} 
and LiB\cite{calandra07}; and ternaries:
CaBeSi,\cite{satta01}
LiBC,\cite{ravindran01,rosner02,dewhurst03,lebegue04,lazicki07}
NaBC,\cite{lebegue04}
MgB$_{2-x}$C$_x$,\cite{medvedeva01prb,medvedeva01jetp}
Mg$_{1-x}$Ca$_x$B$_2$,\cite{medvedeva01prb,medvedeva01jetp,sun07sst,sun07apl}
Mg$_{1-x}$Li$_x$B$_2$,\cite{medvedeva01prb,medvedeva01jetp}
Mg$_{1-x}$Na$_x$B$_2$,\cite{medvedeva01prb,medvedeva01jetp}
MgBe$_x$B$_{2-x}$,\cite{mehl01,medvedeva01jetp} 
CuB$_{2-x}$C$_x$,\cite{mehl01}
MgB$_2$C$_2$,\cite{ravindran01}
and Be$_2$B$_x$C$_{1-x}$.\cite{moussa08}
For these compounds, their structural, electronic, and superconducting properties 
are studied experimetally\cite{sun07sst,sun07apl,rosner01,singh07,gasparov01,lazicki07}
and theoretically.\cite{satta01,medvedeva01prb,medvedeva01jetp,oguchi02,lebegue04,rosner01,mehl01,ravindran01,lazicki07,calandra07,okatov01,moussa08,rosner02,dewhurst03}
The structural stabilities of the existing and hypothetical compounds are studied 
extensively as well.\cite{oguchi02,kolmogorov06}
Even with these efforts, however, the highest $T_c$ observed experimentally 
in these compounds 
is still that of MgB$_2$. There have been theoretical predictions  of higher $T_c$'s than 39~K,
for example, $T_c$ $\sim$ 50~K in CuB$_{2-x}$C$_x$\cite{mehl01} and
$T_c$ $\sim$ 100~K\cite{rosner02} and 65~K\cite{dewhurst03} in hole-doped LiBC.

One of the interesting candidates for a higher $T_c$ is the hypothetical simple hexagonal CaB$_2$,
with the same crystal symmetry as MgB$_2$, with all Mg atoms being replaced by Ca atoms.
First-principles calculations predict that 
the simple hexagonal CaB$_2$ should have a greater unit 
cell (both $a$ and $c$ lattice constants
are longer) and a much larger value for 
the density of states (DOS) at the Fermi energy ($E_F$) than 
MgB$_2$.\cite{medvedeva01jetp,okatov01,medvedeva01prb,ravindran01,oguchi02}
Both features are favorable for a higher $T_c$,
since MgB$_2$ is found to have a positive dependence of $T_c$ on the unit-cell volume\cite{buzea01}
and more generally, a larger DOS at $E_F$ can result in a greater electron-phonon coupling 
constant. 
Although CaB$_2$ is an attractive candidate for superconductivity, 
it has not been synthesized as yet,
nor has it been extensively studied theoretically for superconductivity.
Since the Fermi surface of CaB$_2$ is predicted to consist of
multiple sheets similar to those 
in MgB$_2$,\cite{medvedeva01jetp,okatov01,medvedeva01prb,ravindran01}
it will be necessary to use the anisotropic Eliashberg theory 
to deal with the effects of possible variations of the superconducting energy gap 
on the Fermi surface as in the case of 
MgB$_2$.\cite{choi02nature,choi02prb}

In this work, we perform density-functional {\em ab-initio} pseudopotential calculations
to determine the lattice constants of the hypothetical simple hexagonal phase of CaB$_2$,
and we calculate its electronic structure, phonon spectrum,
and electron-phonon properties. From these material properties,
we construct the fully anisotropic Eliashberg equations to obtain the 
superconducting properties such as $T_c$,
the momentum-dependent superconducting energy gap $\Delta({\bf k},T)$,
and the specific heat as functions of temperature $T$.
These results predict that CaB$_2$ has a higher $T_c$ than MgB$_2$ and a larger difference
between the superconducting energy gap in the $\sigma$ and $\pi$ bands
when compared with MgB$_2$.

\begin{figure} 
\epsfig{file=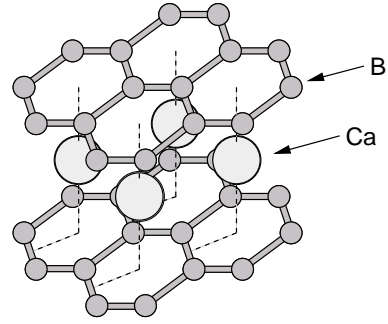,width=5cm,clip=}
\caption{Atomic structure of hypothetical simple hexagonal CaB$_2$. 
The simple hexagonal unit cell contains one Ca and two B atoms.
}
\end{figure}

We consider the hypothetical simple hexagonal CaB$_2$ shown in Fig.~1.
The atomic structure has exactly the same symmetry as MgB$_2$, having one Ca and two B atoms
in a unit cell. We determine the lattice constants $a$ and $c$ by minimizing the total energy
of the system obtained from first-principles pseudopotential density-functional 
calculations.\cite{cohen_t1,ihm} In the calculations, plane waves of energy up to 60 Ry are used to
expand the electronic wavefunctions, norm-conserving pseudopotentials\cite{troullier} are used
to describe the electron-ion interaction, and the local density approximation is employed to deal with
the electron-electron interactions effectively. In more detail, we use the partial-core correction\cite{louie}
for the Ca pseudopotential, Ceperley-Alder exchange-correlation 
energy\cite{ceperley} of
the Perdew-Zunger parameterization,\cite{perdew} a $12 \times 12 \times 12$ $k$-point grid 
to evaluate the self-consistent electron density, and a $18 \times 18 \times 12$ $k$-point grid 
for Fermi-surface properties.

For simple hexgonal CaB$_2$, 
the calculated lattice constants at ambient pressure 
are $a$ = 3.185 {\AA} and $c$ = 4.060 {\AA}. Thus the $c/a$ ratio is 1.27.
Compared with theoretical values for MgB$_2$ (which are $a$ = 3.07 {\AA} and $c$ = 3.57 {\AA},
with $c/a$ = 1.16), $a$ in CaB$_2$ is 4\% greater and $c$ is 14\% greater. 
To check the ratio $c/a$ at high pressure, we obtain the lattice constants 
at 30 GPa, which are $a$ = 3.014 {\AA} and $c$ = 3.842 {\AA}. Thus, the $c/a$ ratio
is still 1.27 at this pressure.
The obtained $c/a$ = 1.27 is larger than 
1.165 known as the maximal $c/a$ ratio of existing diborides,\cite{aronsson65}
and this is in agreement with previous calculations.\cite{medvedeva01jetp,okatov01,medvedeva01prb,ravindran01,oguchi02} 
Since the hexagonal CaB$_2$ has not been synthesized successfully, its formation energy
and stability have been studied theoretically.\cite{medvedeva01prb,oguchi02,kolmogorov06}
Hexagonal CaB$_2$ is predicted to be stable with respect to hcp Ca and rhombohedral 
B ($\alpha$-B$_{12}$),\cite{medvedeva01prb} but unstable 
to a phase separation into fcc Ca and  CaB$_6$,\cite{oguchi02} 
and less stable than $\delta$-CaB$_2$ phase.\cite{kolmogorov06}
As reported theoretically, hexagonal CaB$_2$ may not be the lowest phase energetically, but
our frozen-phonon calculations at high symmetry
points, which will be presented below, do not show any
structural instability from the simple hexagonal phase.

\begin{figure} 
\epsfig{file=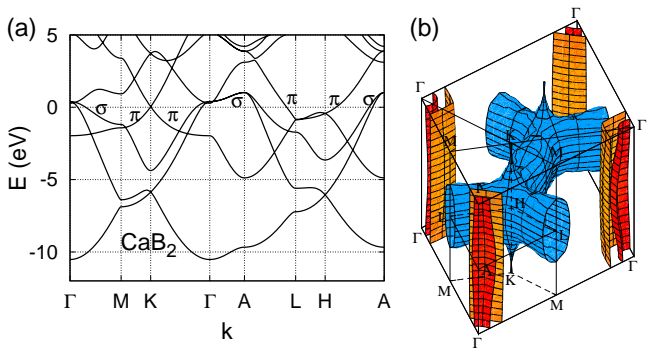,width=8.5cm,angle=0,clip=} 
\caption{(Color online.) Electronic structure of CaB$_2$:
(a) electronic band structure and (b) the Fermi surface.
In (b), the cylindrical (red and orange) sheets 
along the $\Gamma-$A$-\Gamma$ line
are of hole type, and the connected (blue) sheet along H--L lines 
is of electron type.
}
\end{figure}

With the optimized lattice constants, we perform first-principles 
electronic structure calculations. Figure~2 shows the obtained band structure
along the high symmetry lines and the Fermi surface. 
Overall, the band structure of CaB$_2$ is similar to MgB$_2$. 
In Fig.~2(a), the $\sigma$ and the $\pi$ bands are from the boron
$sp^2$ and $p_z$ orbitals, respectively.
The full bandwidth of the $\sigma$ bands is reduced substantially, 
compared with MgB$_2$, because of the larger B--B bond length. 
As in the case of MgB$_2$, the Fermi energy in CaB$_2$
is lower than the top of the $\sigma$ bands,
producing two cylindrical hole-type sheets
along the $\Gamma$--A line [red and orange regions in Fig.~2(b)].
While the hole-type sheets are close to those
in MgB$_2$ qualitatively,
the electron-type sheet from the $\pi$ bands in CaB$_2$ 
is quite different from
the one in MgB$_2$. In MgB$_2$, the $\pi$ bands are above $E_F$ 
at the $M$ and $K$ 
points; however, in CaB$_2$, they are below $E_F$. Thus,
CaB$_2$ has only one electron-type Fermi surface from the $\pi$ bands
[the blue sheet in Fig.~2(b)], 
missing the Fermi surface along the $K$--$L$ line that is present in MgB$_2$.
Hence, as shown in Fig.~2(b),
the Fermi surface in CaB$_2$
consists of three sheets: two hole-type cylindral ones
along $\Gamma$--A lines and one electron-type along H--L lines. 
The density of states at the Fermi level in CaB$_2$ 
is 0.96 states/eV per formula unit, which is much larger than 0.69 states/eV for MgB$_2$. 
The contributions of the three sheets of the Fermi surface
to the DOS at $E_F$ are 12\%, 32\%, and 56\%
from the inner and outer hole-type sheets and
the electron-type sheet, respectively.
Our results for the band structures 
are in good agreement with the results using 
the full-potential linear muffin-tin orbital (FP-LMTO) method.\cite{medvedeva01prb,medvedeva01jetp}

\begin{figure} 
\epsfig{file=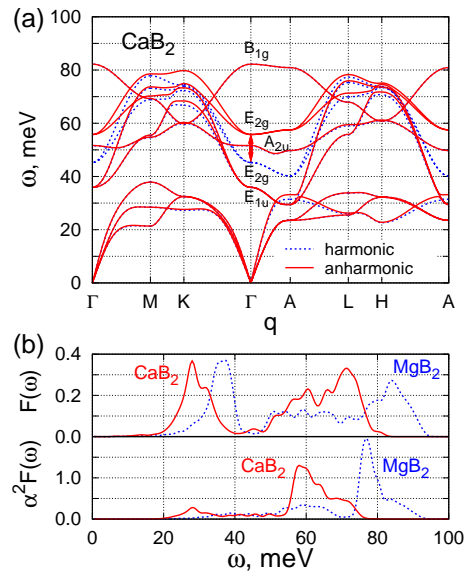,width=6cm,angle=0.0,clip=}
\caption{(Color online.) Phonon dispersion relations and electron-phonon interactions in CaB$_2$.
(a) Dashed (blue) lines are the phonon dispersions within the harmonic 
approximation  and solid (red) lines are those considering 
anharmonicity in the frozen-phonon
calculations. As in the case of MgB$_2$, the doubly degenerate $E_{2g}$ 
mode along the $\Gamma-A$ line is highly affected by anharmonicity.
In the harmonic case, the frequency of the $E_{2g}$ mode is even lower than 
the $A_{2u}$ mode, but the former becomes higher than the latter in the 
anharmonic case. (b) Phonon density of states $F(\omega)$ in meV$^{-1}$
and the Eliashberg function $\alpha^2F(\omega)$ including anharmonic
effects on the phonon frequencies.
Solids (red) lines are 
the phonon density of states and the Eliashberg function in CaB$_2$. Dashed
(blue) lines are those in MgB$_2$ drawn for comparison.
}
\end{figure}

To consider electron-phonon interactions, we calculate the 
phonon structures in CaB$_2$ based on the frozen-phonon method.
In the first step, 
phonon frequencies and polarizations are obtained at high symmetry points
of the Brillouin zone by frozen-phonon methods, and then
interpolated to the full Brillouin zone by interpolating the dynamical
matrix. In the frozen-phonon calculations, we extract
quadratic and higher-order dependences of
the total energy on atomic displacements, and calculate 
harmonic and anharmonic phonon frequencies 
with and without using the harmonic approximation.

Figure~3(a) shows the calculated phonon dispersions along the high symmetry lines,
with and without anharmonicity. The phonon frequencies are
calculated by constructing the dynamical matrices
at the high-symmetry points and interpolating them
along the high-symmetry lines. In our frozen-phonon calculations,
the increase of the total energy with boron 
displacements are highly anharmonic for $E_{2g}$ modes
(the B--B bond stretching modes) at the $\Gamma$ and $A$ points. This is
consistent with the reported large
anharmonicity at $\Gamma$ in CaB$_2$.\cite{yildirim01} 
We find that the $E_{2g}$ frequency is 45.2 meV in the harmonic
case and 55.8 meV in the anharmonic case, both of which are substantially
lower than those in MgB$_2$.
All the phonon frequencies at the high symmetry points
are positive, and therefore do not indicate structural instabilities.

The upper panel in Fig.~3(b) shows the phonon density of states
obtained by interpolating the dynamical matrix
throughout the Brillouin zone with anharmonicity included.
Compared with MgB$_2$, the phonons in CaB$_2$ have lower frequencies.
Phonon modes associated with Ca atoms should have lower frequencies
because of heavier atomic mass of Ca relative to Mg. In addition, phonon 
modes associated with B atoms also have lower frequencies 
because of the elongation of B--B bonds.

We calculate the momentum-dependent Eliashberg function
$\alpha^2F({\bf k},{\bf k'},\omega)$ from the difference in the
self-consistent potential with and without frozen phonons,
using the anharmonic phonon frequencies,
and then take the average on the Fermi surface to obtain
the isotropic Eliashberg function $\alpha^2F(\omega)$,
as shown in the lower panel in Fig.~3(b).
From $\alpha^2F(\omega)$, we can evaluate the average 
electron-phonon coupling constant 
$\lambda = 2\int d\omega \alpha^2F(\omega)/\omega$ 
and the logarithmically averaged phonon frequency
$\omega_{ln}=\exp[(2/\lambda)\int d\omega \alpha^2F(\omega)\ln \omega/\omega]$.
This gives $\lambda$ = 0.69 and 
$\omega_{ln}$ = 50 meV, respectively. 
The value of $\lambda$ is 13\% larger 
and $\omega_{ln}$ is 34\% smaller than those for MgB$_2$.

We obtain the superconducting properties in CaB$_2$ using the 
fully anisotropic Eliashberg formalism.\cite{choi02nature,allen}
The anisotropic Eliashberg equations at imaginary frequencies
are given as
\begin{eqnarray}
\nonumber
Z({\bf k},i\omega_n)  &=& 1  + f_ns_n\!\sum_{{\bf k}'n'}\!
W_{{\bf k}'}\lambda({\bf k},{\bf k}'\!,n\!-\!n')\\
&\times&
\frac{\omega_{n'}}
{\sqrt{\!\omega_{n'}^2\!+\!\Delta\!({\bf k}'\!,i\omega_{n'}\!)^2}}, 
\label{aniso_Z_iw}
\end{eqnarray}
\begin{eqnarray}
\nonumber
Z({\bf k},i\omega_n\!)\Delta({\bf k},i\omega_n\!) &=& \pi T\! 
\sum_{{\bf k}'n'}\!W_{{\bf k}'}\!\left[\lambda({\bf k},\!{\bf k}'\!,\!n\!-\!n')
\!-\!\mu^*\!(\omega_c)\right]\\
&\times&
\frac{\Delta({\bf k}'\!,i\omega_{n'})}{\sqrt{\!\omega_{n'}^2
\!+\!\Delta({\bf k}'\!,i\omega_{n'}\!)^2}},\label{aniso_ZD_iw}
\end{eqnarray}
where $\omega_n = (2n+1)\pi T$ is the fermionic Matsubara frequency at temperature $T$,
$Z({\bf k},i\omega_n)$ and $\Delta({\bf k},i\omega_n)$ are 
the momentum-dependent renormalization function and the 
gap function, respectively, and $\lambda({\bf k},{\bf k'},n)$ represents the 
momentum-dependent electron-phonon interaction. Definitions of symbols and
details of the numerical method are described in 
Refs.~\onlinecite{choi02prb},~\onlinecite{choi03}, and
\onlinecite{choi06}. 
Here, we assume that the Coulomb pseudopotential $\mu^*$ 
is isotropic, since its momentum dependence
is not as strong as that of the electron-phonon interaction,
\cite{moon,mazin,choi04reply}
and assume $\mu^*(\omega_c)$ = 0.12 
for the cut-off frequency $\omega_c$ = 0.5 eV, 
as assumed for MgB$_2$.\cite{choi02nature,choi02prb}

We obtain $T_c$ by finding the highest $T$ at which
the anisotropic Eliashberg equations [Eqs.~(1) and (2)] have 
a non-trivial solution for $\Delta({\bf k})$.
With the assumed value of $\mu^*(\omega_c)$ = 0.12, 
we obtain $T_c$ = 48 K, which is about 10 K higher
than that of MgB$_2$. To check the sensitivity of $T_c$ to $\mu^*$, we also
consider $\mu^*(\omega_c)$ = 0.10 and 0.14, obtaining 
$T_c$ = 50 and 46 K, respectively. Thus,
we can expect a higher $T_c$ in CaB$_2$ than in MgB$_2$ 
even in the case of relative large $\mu^*$.

\begin{figure} 
\epsfig{file=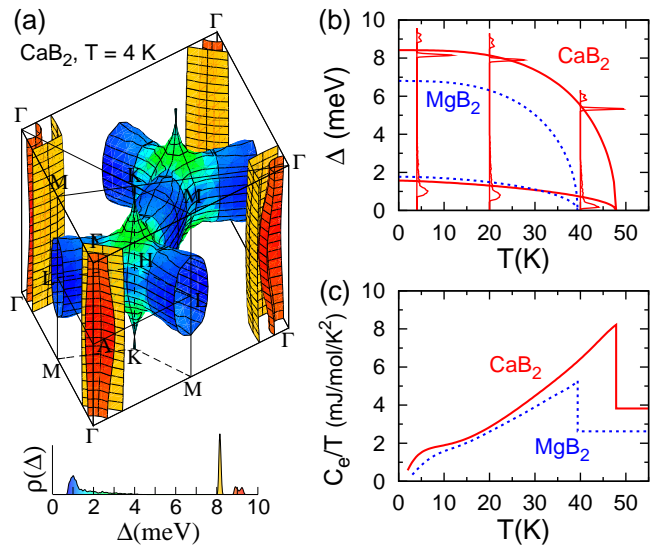,width=8.5cm,angle=0,clip=} 
\caption{(Color online.) Superconducting properties in CaB$_2$.
(a) Superconducting energy gap $\Delta({\bf k})$ on the Fermi surface at 4 K
plotted using a color scale. The lower panel shows the distribution $\rho(\Delta)$ of the
values of the energy gap as well as the color scale for the size of the energy gap.
(b) Temperature dependence of the superconducting energy gaps
 [solid (red) lines].  Dashed (blue) lines are the superconducting 
energy gaps in MgB$_2$ drawn for comparision.
Vertical lines represent the distributions of $\Delta({\bf k})$ in CaB$_2$
at 4, 20, and 40 K. Curves are fit to the separate averages, $\Delta_{\sigma}$ and $\Delta_{\pi}$,
where the average $\Delta_{\pi}$ is greater than the peak energy of $\rho(\Delta)$ in (a) because
of the high-energy tail of $\rho(\Delta)$ [green part in (a)].
(c) The specific heat $C_e$ over $T$ of CaB$_2$ [solid (red) line].
 The dashed (blue) line is $C_e/T$ of MgB$_2$ drawn for comparision.
Anharmonic effects on the phonon frequencies are considered in all
calculations of the superconducting properties.
}
\end{figure}

With Eqs.~(1) and (2), and their analytic continuations to 
the real-frequency axis, we calculate the superconducting
properties below $T_c$.
Figure~4(a) shows the calculated superconducting 
energy gap on the Fermi surface at $T$ $=$ 4 K,
where the average values of $\Delta({\bf k}$) on the 
three sheets of the Fermi surface are well separated from one another 
and they are 9.1, 8.2, 
and 1.5 meV, respectively, with the average on the two hole-type 
sheets being 8.4 meV. 
As shown in Fig. 4(b), 
when compared with MgB$_2$, 
the average value $\Delta_\sigma$ of $\Delta({\bf k})$ 
on the $\sigma$ bands (the hole-type sheets) 
is substantially larger but $\Delta_\pi$ for the $\pi$ bands
is slightly smaller at low $T$, indicating an enhanced two-gap 
nature for CaB$_2$.
Figure~4(c) shows the calculated electronic specific heat as a function
of temperature.
Because of the higher DOS at $E_F$ and the larger $\lambda$, the normal-state
specific heat $C_N$ above $T_c$ is expected to be larger in CaB$_2$
($C_N$ = $\gamma_nT$ with $\gamma_n$ = 3.82 mJmol$^{-1}$K$^{-2}$)
than in MgB$_2$ ($\gamma_n$ = 2.62 mJmol$^{-1}$K$^{-2}$).
In addition, since $\Delta_\pi$ in CaB$_2$ is smaller in magnitude than in MgB$_2$,
the specific heat at $T$ $<$ 10 K has a bigger hump in CaB$_2$.

To summarize, we obtained the lattice constants, the electronic 
and phononic structures, and electron-phonon interactions in 
hypothetical simple hexagonal CaB$_2$ using first-principles 
calculations. We then calculated the superconducting transition 
temperature $T_c$, the superconducting energy gap $\Delta({\bf k},T)$, 
and the specific heat using the fully anisotropic Eliashberg 
formalism. The obtained Fermi surface in CaB$_2$ consists of 
three sheets rather than four as in MgB$_2$, and the phonon frequencies are lower than in 
MgB$_2$. A substantially higher $T_c$ is predicted for 
hexagonal CaB$_2$, and furthermore the multiple superconducting 
energy-gap feature is expected to be enhanced in CaB$_2$, 
with larger $\Delta_\sigma$ 
and smaller $\Delta_\pi$ than in MgB$_2$.

This work was supported by the KRF (Grant No. KRF-2007-314-C00075),
by the KOSEF under Grant No. R01-2007-000-20922-0, 
by NSF under Grant No. DMR07-05941, and 
by the Director, Office of Science, Office of Basic Energy Sciences, 
Materials Sciences and Engineering Division, U.S. Department of
Energy under Contract No. DE-AC02-05CH11231. 
Computational resources have been 
provided by KISTI Supercomputing Center (Project No. KSC-2008-S02-0004),
NSF through TeraGrid resources at SDSC, and DOE at Lawrence Berkeley
National Laboratory's NERSC facility.

\end{document}